# Fast Energy-Theft Attack on Frequency-Varying Wireless Power without Additional Sensors

Hui Wang, Nima Tashakor, Xiaoyang Tian, Hans D. Schotten, and Stefan M. Goetz

*Abstract*—With the popularity of wireless charging, energy access protection and cybersecurity are gaining importance, especially in public places. Currently, the most common energy encryption method uses frequency and associated impedance variation. However, we have proven that this method is not reliable, since a hacker can detect the changing frequency and adjust the compensation. However, the previously presented system needed time to follow the updated frequency, while encryption systems may vary the frequency faster to avoid energy theft. Furthermore, the previous system required an additional sensor coil. To solve these problems, we optimized the attack and the associated system, which can intrude and steal energy within 200 µs. The key is the elimination of the time-consuming maximum receiver current regulation. Also, we use the main receiving coil rather than any additional sensor antenna to detect the magnetic field. Thus, the new hardware is even simpler. A simulation model and experimental results demonstrate the fast response speed of the attack on encrypted wireless power and steal 65% of the power. Overall, the applicability of the attack is highly improved and leaves less room for hardening the encryption. The results demonstrate that energy access protection needs to be given great attention.

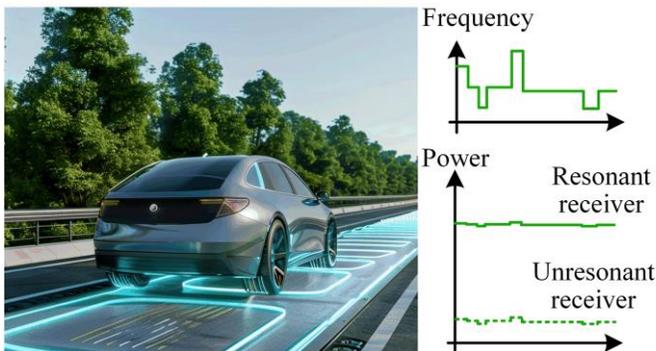

Fig. 1. Frequency-varying energy encryption methods for WPT.

## I. Introduction

OVER the past decades, more and more electronic devices implement flexible wireless charging [1, 2]. Not only because wireless power transfer (WPT) has become more reliable, safe, and efficient than before [3, 4], but also because it is very flexible and even can charge moving devices [5]. Thus, WPT has been widely used for implants [6, 7], electric motors [8, 9], medical devices [10], electric vehicles (EVs) [11], and smartphones [12]. Especially, many EV companies and governments consider wireless charging as an important means to make up for insufficient battery performance, such as Sweden, which is building one of the world's first electrified roads for EVs [13].

In fact, many public spaces have charging ports, wireless ones so far typically for smaller devices. Some of them are public welfare for free, while others are privately run and only provide paid services. However, cost recovery is a serious problem for these paid services, particularly for EV roadway charging [14]. Any EV exposed to this magnetic field could have the chance to use this wireless energy, whether it pays or not [15]. Unauthorized energy hackers can seriously impact the safety and reliability of dynamic road charging infrastructure for EVs:

1) First, unauthorized users can steal power, which causes financial losses to operators as well as their funders and undermines the business model of charging services. Since there may be high numbers of EVs in the future, even if only a small share of them would steal energy, it can risk the entire investment case.

2) Second, unauthorized energy harvesting can cause power fluctuations that affect the reliability of charging services for authorized users.

3) Third, if users believe that the system is unsafe or unreliable due to unauthorized energy harvesting, it may lead to a decrease in usage and trust in the charging infrastructure.

To keep energy safe, multiple energy encryption methods have been proposed. Magnetic field editing shapes the magnetic field spatially and lets it extend to only a small range, whereas unauthorized users do not get physical access [16, 17]. However, this strategy requires the transmitter to acknowledge the subscribed user's precise position, and it is particularly difficult for moving users, such as dynamic roadway charging.

For dynamic charging, frequency-varying encryption appears as the most mature method (Fig. 1) [18]. From the perspective of an observer, the transmitter randomly changes its frequency and effective source impedance in a random sequence and is therefore related to stream ciphers for data [19, 20]. Based on the most elementary LC (inductor–capacitor) resonant characteristics, only the resonant receiver can harvest energy efficiently, while nonresonant receivers suffer from high impedance [21]. Thus, the WPT frequency (sequence) can be interpreted as a key for this WPT network, and multiple compensations for frequency-varying systems have been proposed

[22-26]. Moreover, some smart systems keep detecting the difference between transmitted power and the power received by authorized users. Once this difference is unreasonably large, the system will know some unauthorized users are also resonant at the current frequency and steal energy; it can abandon this frequency point to avoid energy theft, but may also terminate authorized users [27].

We have previously demonstrated that the frequency-varying or -hopping strategy is unreliable. We could decipher the key (frequency sequence) with a magnetic sensor coil and fully compensate the transmitter with a time-varying switched-capacitor circuit at a wide frequency range [28]. Even though the energy encryption system may detect energy loss, the system cannot abandon all frequency points. However, our previous demonstrator had limited response speed and needed time in the millisecond range for detection and synchronization to a new frequency. Thus, some readers may mistakenly believe that the frequency-varying encryption method is still safe as long as the frequency hops rapidly.

To further prove that simple frequency-hopping is unreliable, we optimized this attack and the setup to perform it so that the system can hack the wireless energy within hundreds of microseconds. For comparison, the single-trip communication latencies and synchronization accuracy of short-link radio, such as Bluetooth (IEEE 802.15), typically exceeds 5 ms, of WiFi (IEEE 802.11) in the best case still 3 ms [29, 30], and of satellite transmission as now popular in remote areas even for low-earth installations dozens of milliseconds [31]. These timing capabilities are important as they can be used to concurrently exchange the key sequence and synchronize an authorized receiver. Even though the 3rd Generation Partnership Project (3GPP) enabled an ultra-reliable low-latency channel for fifth-generation communication technology (5G) of less than one millisecond [32], this mode is not designed for very high numbers of participants, and normal users will not have access to it to avoid congestion. Regular 5G may stay above 10 ms latency without guaranteed real-time conditions [33].

Thus, our new method has good chances of hacking the system even faster than authorized users can follow and synchronize to the sequence. Also, a driving car normally needs on the order of 30 ms to pass a one-meter diameter coil, so unsubscribed users would have enough time to harvest energy from each small transmitting coil with our exploit. The key to fast energy encryption is that we improved the regulation method for the duty cycle of the switched-capacitor array so that the initial duty cycle setting is already precise and does not need a time-consuming receiver current measurement. Moreover, we detect the WPT frequency and phase through the receiver coil and the compensation capacitor directly, rather than any additional magnetic sensor coil. Thus, the size of the hacking setup is smaller, and the system is simpler.

This paper is organized as follows: Section II presents the system configuration. Then, Section III describes the system operation and design procedures. Sections IV and V respectively demonstrate the system with simulations and experiments. Finally, Section VI summarizes the paper.

## II. System Configuration

Figure 2 presents the topology of a frequency-varying system with authorized users, unauthorized users, and our new hacking device, where $L_T$ and $L_R$ respectively are the transmitter and hacking receiver; $C_{R1}$, $C_{R2}$, and $S_R$ compose the compensation network of the hacking circuit; $I_T$ and $I_R$ respectively are currents of transmitter and hacking receiver; $M_R$ denotes the mutual inductance between transmitter and receiver; $V_{CR1}$, $V_{CR2}$, $V_{SR}$, $I_{R1}$, and $I_{R2}$ are the voltages and currents of corresponding capacitors and the switch; $V_{LR}$ and $V_{RL}$ respectively are the voltages of $L_R$ and the load $R_L$.

As in previous work on unauthorized wireless power theft, we do not know the specific designation of the transmitting and receiving circuits, but we do assume the transmitter $L_T$ can provide wireless energy at a wide frequency range [18]. For the hacking side, the receiver $L_R$ still needs $C_{R1}$ and $C_{R2}$ for compensation, and the switch $S_R$ determines the equivalent capacitance of the switched-capacitor array $C_{RE}$ by controlling the duty cycle of $C_{R2}$. However, in contrast to previous attacks, the micro-controller unit (MCU) acquires the frequency and phase from the capacitor or the load rather than any additional magnetic sensor coil.

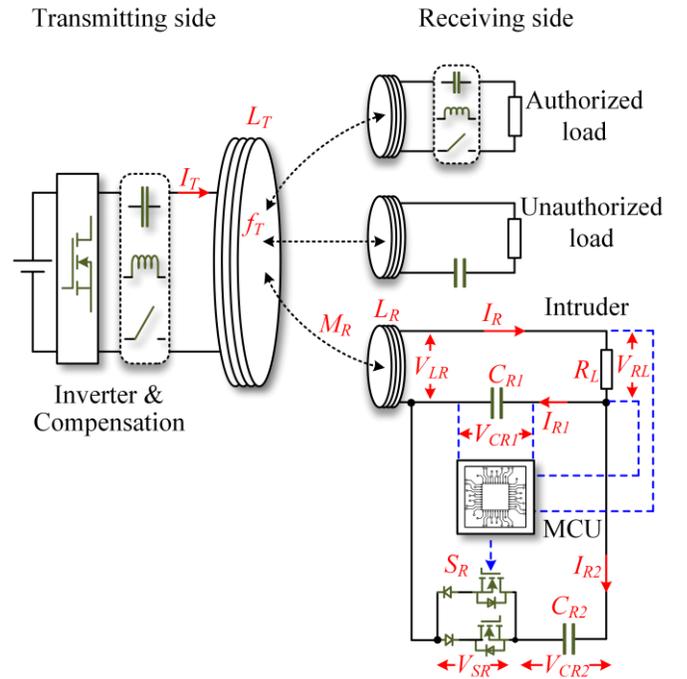

Fig. 2. Common wireless charging system with multi-user and the proposed hacking receiver.

## III. Hacking System Design and Operation

### A. Compensation Selection

Research has studied various compensation networks for multi-frequency WPT (Table I). After weighing the pros and cons of all these compensation networks, we chose a time-division switched-capacitor array because of its simple structure and its ability to be resonant over a wide frequency range. Even though this method requires accurate detection of phase information in real time, we propose an easy solution to this problem easily (detailed in Section III, Part B).

TABLE I
COMPARISON OF MULTI-FREQUENCY COMPENSATIONS

| Compensation | Advantage | Disadvantage |
|---|---|---|
| Switched-capacitor matrix or array [19] | ▪ Multiple resonant frequency points<br>▪ No phase detection<br>▪ Suitable for high power | ▪ A huge number of capacitors and switches |
| High-order network [34-36] | ▪ Multiple resonant frequency points<br>▪ No phase detection<br>▪ Suitable for high power | ▪ Limited resonant frequency points<br>▪ Very complex |
| Variable capacitor [37] | ▪ Wide frequency range<br>▪ Simple and small<br>▪ No phase detection | ▪ Not suitable for high power |
| Time-division switched capacitor [28] | ▪ Wide frequency range<br>▪ Simple and small<br>▪ Suitable for high power | ▪ Need phase information |

### B. Frequency and Phase Detection

The previous system used an additional magnetic flux pickup coil to detect the magnetic field frequency $f_T$ and phase. Even though any coil—including the hacking receiver $L_R$—can induce the corresponding voltage, we did not use $L_R$ because the receiver current $I_R$ may affect the voltage $V_{LR}$, as $I_R$ can be expressed as [38]

$$I_R = \frac{j2\pi f_T M_R I_T}{j2\pi f_R L_R + \frac{1}{j2\pi f_R C_{RE}} + R_L}. \quad (1)$$

When the hacker controls $S_R$ to a wrong frequency $f_R$, the equivalent capacitance $C_{RE}$ irregularly varies from $C_{R1}$ to $C_{R1}+C_{R2}$. Thus, the current $I_R$ is distorted, and the voltage of the coil, capacitor, as well as load becomes [39]

$$\begin{cases} V_{LR} = j2\pi f_T M_R I_T - j2\pi f_R L_R I_R, \\ V_{CR1} = \dfrac{I_R}{j2\pi f_R C_{RE}}, \\ V_{RL} = R_L I_R. \end{cases} \quad (2)$$

Thus, neither the voltages $V_{LR}$, $V_{CR1}$, $V_{RL}$ nor the current $I_R$ can provide a useful magnetic field signal for the controller. Therefore, an open circuit coil without any current is promised to be a reasonable solution for the magnetic sensor in the state of the art. It can provide accurate frequency and phase information as the Kelvin-connected field detection [40]. By timing several upward zero crossings, the controller can easily measure the period and calculate the frequency $f_T$.

However, a large separate sensor coil increases the size and more importantly the cost of the system, whereas a small sensor may miss some magnetic fields. Especially, if the charging coils are buried underground without any sign, an unsubscribed EV user with the intention to steal energy may need a big sensor coil or multiple small sensor coils to find the magnetic field. Either choice will complicate the system. Thus, a large power-receiving coil should ideally perform both energy harvesting and magnetic field detection. One of the key contributions of this paper is the avoidance of the aforementioned receiving current $I_R$ interference.

In the first step, specifically the initial phase detection, the switch $S_R$ is off. Thus, only capacitor $C_{R1}$ compensates $L_R$. Even though the receiving circuit is most likely nonresonant under current conditions, the impedance of the hacking circuit is fixed so that (1) becomes

$$I_R = \frac{j2\pi f_T M_R}{j2\pi f_T L_R + \dfrac{1}{j2\pi f_T C_{R1}} + R_L} \times I_T. \quad (3)$$

Thus, the current $I_R$ should be a sinusoidal waveform with the same frequency as $I_T$, only the phase may be different. Then, (2) becomes

$$\begin{cases} V_{LR} = j2\pi f_T M_R I_T - j2\pi f_T L_R I_R, \\ V_{CR1} = \dfrac{I_R}{j2\pi f_T C_{R1}} \propto I_T, \\ V_{RL} = R_L I_R \propto I_T. \end{cases} \quad (4)$$

To be more specific, the aforementioned waveforms are shown in Fig. 3.

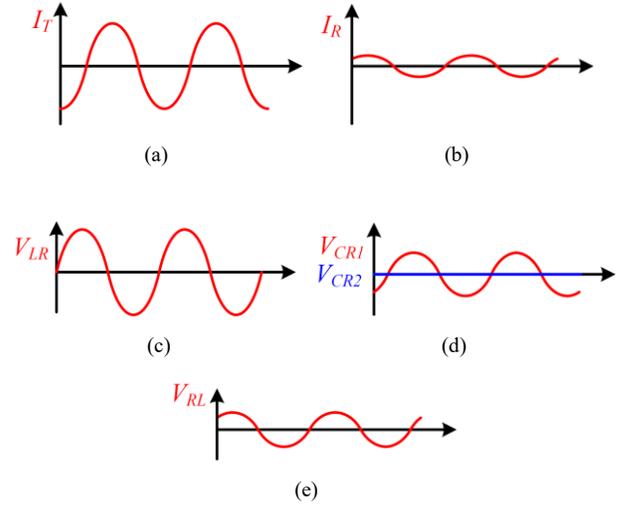

Fig. 3. Waveforms of system voltage and current under unresonant conditions. (a) Transmitter current. (b) Receiver current. (c) Receiver voltage. (d) The voltage of capacitors. (e) The load voltage.

It is clear that all $V_{LR}$, $V_{CR1}$, and $V_{RL}$ can provide magnetic field information directly to the controller. However, the phase difference between $V_{LR}$ and $V_{CR1}$ is uncertain, which highly depends on the mutual inductance $M_R$, load impedance $R_L$, and frequency $f_T$. While it is not easy to control $C_{R2}$ with $V_{LR}$, the voltages of capacitor $V_{CR1}$ and the load $V_{RL}$ offer the chance to replace the sensor coil in magnetic field detection [28].

It should be mentioned that the impedance of the load $R_L$ is generally much smaller than $C_{R1}$ at high frequencies. Thus, the voltage of $V_{CR1}$ should be very small, especially with negligible

$I_R$ under nonresonant conditions. Also, once the load is not purely resistive, the phase difference between $V_{RL}$ and $V_{CR1}$ should vary with the frequency.

Therefore, with a small or not purely resistive load, we prefer to use $V_{CR1}$ as the signal. When the load is large and purely resistive, however, it might be better to extract the signal from $V_{RL}$. The main reason is that signal extraction generally requires filtering. So, the filter time $T_{\text{filter}}$ will shorten the control range of duty cycle $T_{\text{ON}}$, as shown in Fig. 4(a). However, with $V_{RL}$, the controller can acquire the zero phase of $V_{CR1}$ a quarter of a cycle earlier, which means we can filter $V_{RL}$ for a long time and still control the switch $S_R$ in time, as shown in Fig. 4(b).

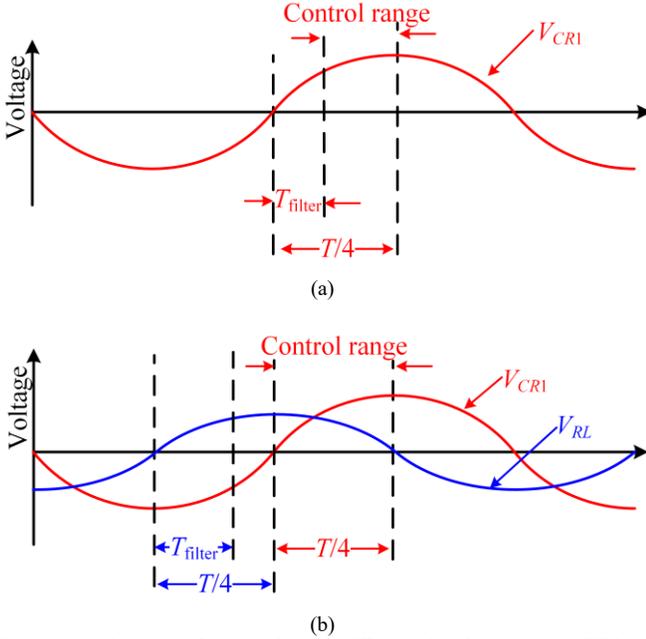

Fig. 4. Control ranges of switch $S_R$ with different signal sources. (a) With capacitor voltage $V_{CR1}$. (b) With load voltage $V_{RL}$.

### C. Fast Frequency Recognition Strategy

In this hacking system, the controller labels and memorizes every frequency that has been hacked because most encryption systems still prefer to use several fixed frequency points for hacking. The fixed frequencies are a consequence of traditional capacitor arrays [22] and matrices [19] with a finite number of resonances. For high-efficiency targets and low unintended electromagnetic emissions, those discrete resonances at present outperform any fully tunable circuit. Thus, when the controller detects the transmitter frequency hopping, the controller will detect the new frequency for a short time and assume the nearest previously encrypted frequency as the first fast frequency detection estimate. Then, the controller will control the switch $S_R$ with the corresponding $T_{\text{ON}}/T_{\text{OFF}}$ times first and keep detecting the new frequency as well as refining the estimate continuously.

### D. Fast and Precise Control of Variable Compensation

The controller compensates the receiver of the intruder with a time-division switched-capacitor array, which controls the duty cycle of capacitor $C_{R2}$ to regulate the equivalent capacitance $C_{RE}$ in each WPT cycle. In contrast to previous work, however, we do not rely on the slow analog-digital conversion (ADC) to tune the switch on/off time $T_{\text{ON}}/T_{\text{OFF}}$, but detect the phase difference between $I_R$ and $I_T$. Also, we calculate and tune $T_{\text{ON}}/T_{\text{OFF}}$ in advance. Since the correct $T_{\text{ON}}/T_{\text{OFF}}$ primarily depends on the frequency and does not relate to the load condition or other variables, the new system can record the ideal $T_{\text{ON}}/T_{\text{OFF}}$ of each frequency point in advance and save hundreds of milliseconds in real applications.

**For the calculation part:** We use the equation from the literature [28] to calculate the switch on/off time per

$$\begin{cases} T_{\text{ON}} = \dfrac{1}{\pi f_T}\arcsin\left(\dfrac{C_{R1}+C_{R2}}{C_{R2}} - (2\pi f_T)^2 L_R \dfrac{C_{R1}(C_{R1}+C_{R2})}{C_{R2}}\right), \\ T_{\text{OFF}} = \dfrac{T}{2} - T_{\text{ON}}. \end{cases} \quad (5)$$

Even though the derivation of this equation is based on a number of approximations, it allows a hacker to steal significant energy without any payment. Also, the precise calculation of $T_{\text{ON}}/T_{\text{OFF}}$ is complicated and includes many unknown variables.

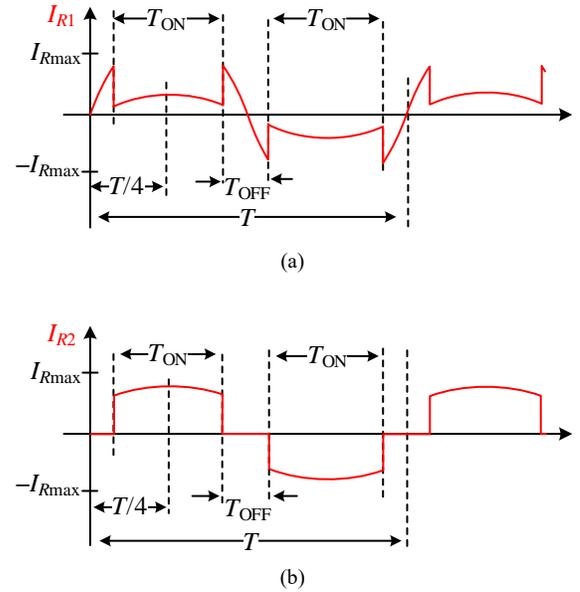

Fig. 5. Capacitor currents. (a) The current of the first capacitor $C_{R1}$. (b) The current of the second capacitor $C_{R2}$.

More specifically, the actual current waveforms of the intruder's compensation (Fig. 5) can be expressed as

$$I_{R1} = \begin{cases} I_{R\max 1}\sin(2\pi f_{R1}t) & \left(t \leq \dfrac{T_{\text{OFF}}}{2}\right) \\ K_1\left[I_{R\max 12}\sin(2\pi f_{R12}t+\alpha_T)+\beta_{T1}\right] & \left(\dfrac{T_{\text{OFF}}}{2} \leq t \leq \dfrac{T}{4}\right) \end{cases} \quad (6)$$

and

$$I_{R2} = \begin{cases} 0 & \left(t \leq \dfrac{T_{\text{OFF}}}{2}\right) \\ K_2\left[I_{R\max 12}\sin(2\pi f_{R12}t+\alpha_T)+\beta_T\right] & \left(\dfrac{T_{\text{OFF}}}{2} \leq t \leq \dfrac{T}{4}\right) \end{cases}. \quad (7)$$

The frequency $f_{R1}$ is between $f_T$ and the resonant frequency of $C_{R1}$ as well as $L_R$, while the frequency $f_{R12}$ is between $f_T$ and the resonant frequency of paralleled $C_{R1}\|C_{R2}$ as well as $L_R$. These are highly related to the switch on/off time. When the intercepting receiver is at the full resonant condition, $f_{R1}$ should be very close to the resonant frequency of $C_{R1}$ and $L_R$, while $f_{R12}$ is near the resonant frequency of paralleled $C_{R1}\|C_{R2}$ and $L_R$. However, the specific values of them are unknown. Additionally, the phase offset $\theta_T$ and amplitude offset $K_T$ are also uncertain. Moreover, $I_{R\max 1}$ and $I_{R\max 12}$ are the current amplitudes, which should be near the full resonant receiver maximum current $I_{R\max}$. Only the current splitting factors of paralleled $C_{R1}\|C_{R2}$, namely, $K_1$ and $K_2$, have approximate values and can be expressed as

$$\begin{cases} K_1 = \left| \dfrac{\dfrac{-j}{j2\pi f_{R12} C_{R2}} + \dfrac{\Delta V_d}{I_{R1}}}{\dfrac{-j}{2\pi f_{R12} C_{R1}} + \dfrac{-j}{2\pi f_{R12} C_{R2}} + \dfrac{\Delta V_d}{I_{R1}}} \right| \approx \dfrac{C_{R1}}{C_{R1}+C_{R2}}, \\ K_2 = \left| \dfrac{\dfrac{-j}{2\pi f_{R12} C_{R1}}}{\dfrac{-j}{2\pi f_{R12} C_{R1}} + \dfrac{-j}{2\pi f_{R12} C_{R2}} + \dfrac{\Delta V_d}{I_{R1}}} \right| \approx \dfrac{C_{R2}}{C_{R1}+C_{R2}}, \end{cases} \quad (8)$$

where $\Delta V_d$ is the forward voltage drop of the transistor and diode at the switch $S_R$.

For practicality, we replace the piecewise equation with a sinusoidal current, which is expressed as

$$I_R = I_{R\max} \sin(2\pi f_T t) \quad \left(0 \le t \le \dfrac{T}{4}\right). \quad (9)$$

Therefore, the current waveforms (6) and (7) are simplified as

$$I_{R1} = \begin{cases} I_{R\max} \sin(2\pi f_T t) & \left(t \le \dfrac{T_{\text{OFF}}}{2}\right) \\ \dfrac{C_{R1}}{C_{R1}+C_{R2}} I_{R\max} \sin(2\pi f_T t) & \left(\dfrac{T_{\text{OFF}}}{2} \le t \le \dfrac{T}{4}\right) \end{cases} \quad (10)$$

and

$$I_{R2} = \begin{cases} 0 & \left(t \le \dfrac{T_{\text{OFF}}}{2}\right) \\ \dfrac{C_{R2}}{C_{R1}+C_{R2}} I_{R\max} \sin(2\pi f_T t) & \left(\dfrac{T_{\text{OFF}}}{2} \le t \le \dfrac{T}{4}\right) \end{cases}. \quad (11)$$

Then, as a substitute for the full resonant capacitor, the adopted switched capacitor array and $C_{RE}$ share similar electrical properties. For instance, they share the same effective capacitor voltage for equal current, which leads to

$$\dfrac{\int_0^{\frac{T}{4}} I_R dt}{C_{RE}} = \dfrac{\int_0^{\frac{T}{4}} I_{R1} dt}{C_{R1}} = V_{SR} + \dfrac{\int_0^{\frac{T}{4}} I_{R2} dt}{C_{R2}}. \quad (12)$$

Even though the switch voltage $V_{SR}$ is unknown, the voltage change of the first capacitor $C_{R1}$ can be easily acquired from

$$V_{CR1} = \dfrac{\int_0^{\frac{T}{4}} I_{R1} dt}{C_{R1}} = \dfrac{\int_0^{\frac{T_{\text{OFF}}}{2}} I_R dt + \dfrac{C_{R1}}{C_{R1}+C_{R2}} \int_{\frac{T_{\text{OFF}}}{2}}^{\frac{T}{4}} I_R dt}{C_{R1}}. \quad (13)$$

Thus, from (12) and (13), we can get

$$C_{RE} = \dfrac{1}{\dfrac{1-\cos(\pi f_T T_{\text{OFF}})}{C_{R1}} + \dfrac{\cos(\pi f_T T_{\text{OFF}})}{C_{R1}+C_{R2}}}. \quad (14)$$

Based on the most basic LC resonant circuit, the ideal equivalent capacitance $C_{RE}$ should be [41]

$$C_{RE} = \dfrac{1}{(2\pi f_T)^2 L_R}. \quad (15)$$

Therefore, from (14) and (15), we can get the aforementioned (5), which can calculate the switch on/off time from several known or measured values, such as $C_{R1}$, $C_{R2}$, $L_R$, and $f_T$.

**For the regulation part:** The system can detect the phase difference between the transmitter current $I_T$ and the receiver current $I_R$. According to (3), when the capacitance of the time-division switched capacitor array $C_{RE}$ is insufficient, $I_R$ leads $I_T$ more than 90 degrees, so the controller can increase $T_{ON}$ to let $C_{R2}$ participate more in the compensation process. However, when $C_{RE}$ is too high, $I_T$ lags behind $I_R$ by less than 90° and the controller can decrease $T_{ON}$ to reduce $C_{RE}$.

A comparator for the detection of the zero-phase point of $I_R$ (or $V_{RL}$) and $I_T$ turns the phase detection into simple digital I/O signal processing that can be done at high speed. In contrast to maximum-current tracking [28], phase adjustment furthermore avoids optimization but is a monotonic process that can use easy and fast control.

### E. System Design

The inductance $L_R$ of the intercepting receiver is preferably chosen as high as possible for a sufficient mutual inductance $M_R$ and a large signal. After winding the receiver $L_R$, the capacitors $C_{R1}$ and $C_{R2}$ should follow

$$\begin{cases} C_{R1} \le \dfrac{1}{(2\pi f_H)^2 L_R}, \\ C_{R2} \ge \dfrac{1}{(2\pi f_L)^2 L_R} - C_{R1}, \end{cases} \quad (16)$$

where $f_H$ and $f_L$ denote the upper and lower limits of the frequency range to be covered for stealing power.

Importantly, (16) is applicable only when the system adopts $V_{RL}$ as the sensor signal. When the system uses $V_{CR1}$ as the sensor signal, a short filter time $T_{\text{filter}}$ restricts the upper limit of the hacking frequency range, as shown in Fig. 4(a). Thus, from (14) and (15), the maximum of $C_{R1}$ can be acquired from

$$\dfrac{1}{(2\pi f_H)^2 L_R} = \dfrac{1}{\dfrac{1-\sin(\pi f_H T_{\text{filter}})}{C_{R1}} + \dfrac{\sin(\pi f_H T_{\text{filter}})}{C_{R1}+C_{R2}}}. \quad (17)$$

The selection of the capacitance should follow

$$\begin{cases} C_{R2} \ge \dfrac{1}{(2\pi f_L)^2 L_R}, \\ C_{R1} \le \dfrac{1-\sin(\pi f_H T_{\text{filter}})}{(2\pi f_H)^2 L_R - \dfrac{\sin(\pi f_H T_{\text{filter}})}{C_{R2}}}. \end{cases} \quad (18)$$

The switch includes two MOSFETs and two diodes (Fig. 2). The diodes and MOSFETs must be able to resist high voltage and suitable for high-frequency operation, at least several hundred thousand Hertz.

*F. Overall Process of the Attack*

Fig. 6 illustrates the flowchart of the system design and operation. After determining the hacking frequency range and winding the receiver coil, we can select the compensation capacitors $C_{R1}$ and $C_{R2}$ through (16) or (18). Then, the hacking receiver calculates and calibrates the $T_{ON}/T_{OFF}$ times of each frequency point first.

In a real-life application, the system would use a large receiver to sense any available magnetic field. Switch $S_R$ is off at the initial stage of sensing. If there is any upward zero-phase point at $V_{RL}$ or $V_{CR1}$, the controller will estimate the frequency for a short interval, such as 100 µs. If the estimated frequency nears any most recently detected frequency within 1 kHz, the controller adopts the previously detected value. Subsequently, the controller actuates the switch $S_R$ with the corresponding $T_{ON}/T_{OFF}$ according to (5). Meanwhile, the controller can progressively refine the frequency detection.

When the capacitor voltage $V_{CR1}$ or load voltage $V_{RL}$ distorts and contains multiple unreasonable zero-crossings, the receiver believes the frequency detection failed, e.g., because the transmitter frequency $f_T$ jumped during the estimation interval. In response, the hacking receiver turns off the switch $S_R$ and restarts the first stage.

Importantly, the controller does not need to switch off $S_R$ every time when the $f_T$ hopped. If the new frequency is near the older one, the former switch-on/off time $T_{ON}/T_{OFF}$ can still achieve stable operation, even though the current $C_{RE}$ cannot compensate the receiver $L_R$ for full resonance. Thus, the system can detect the frequency through $R_L$ or $C_{R1}$ directly.

Compared to the previous hack, the new approach has many unique advantages (listed in Table II), such as being simpler and faster.

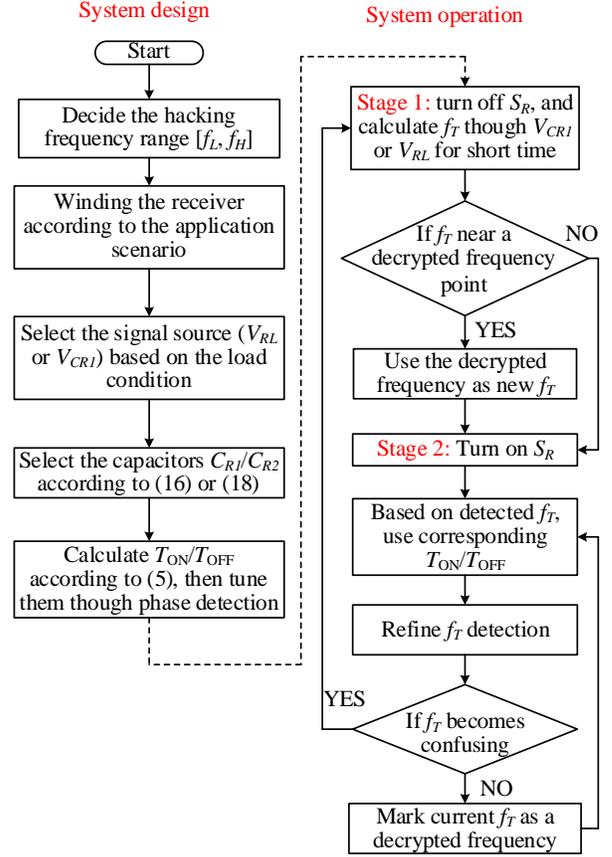

Fig. 6. Flowchart of design and operation of the proposed decryption system.

TABLE II
DIFFERENCE BETWEEN FORMER AND NEW ATTACK

|  | Former system [28] | New system |
|---|---|---|
| Number of magnetic sensor coil | ≥1 | 0 |
| Magnetic field detection area | Same as the sensor coil(s) | Same as the power receiver coil |
| Current sensor | Needed | Not needed |
| Duty cycle regulation | Calculation via (14) & maximum current tracking | Table lookup |
| Hacking preparation time | ~100 ms | <0.2 ms |

## IV. SIMULATION

We analyzed the performance of the method in a Matlab/Simulink model. We additionally implemented the most potent system from the literature for comparison [28]. Furthermore, we picked the parameters of the physical elements similar to the prior art. The transmitter and receiver coils $L_T$ and $L_R$ are respectively 150 µH and 80 µH; the compensation $C_{R1}$ and $C_{R2}$ are respectively 3 nF and 130 nF. The frequency range the receiver could track ranges from 50 kHz to 324 kHz (by detecting $V_{RL}$).

Consistent with the theoretical analysis, the simulation results clearly show the phase difference between the transmitter current and receiver current. When the transmitter $L_T$ offers load-independent current $I_T$ at 300 kHz (Fig. 7(a)), the receiver current $I_{RF}$ from the literature leads $I_T$ for more than a quarter of a cycle. Thus, the new system uses a longer $T_{ON}$ (1.6 times), according to (3). Consequently, the new system receiver current $I_{RN}$ is larger, as $I_{RF}:I_{RN}:I_{R300}$ is 1.32:1.39:1.4, where $I_{R300}$ is the full resonant receiver current at 300 kHz.

When $f_T$ reaches 120 kHz, the former system receiver current $I_{RF}$ leads $I_T$ by less than a quarter of a period, while the full resonant receiver current $I_{R120}$ is exactly 90° ahead of $I_T$. Thus, the new hacking device shortens $T_{ON}$ (to 95% of the calculated value) for lower $C_{RE}$, and $I_{RN}$ shares the same phase with $I_{R120}$, as shown in Fig. 7(b). As a result, the ratio of rated values $I_{RF}:I_{RN}:I_{R120}$ is 1.48:1.64:1.7.

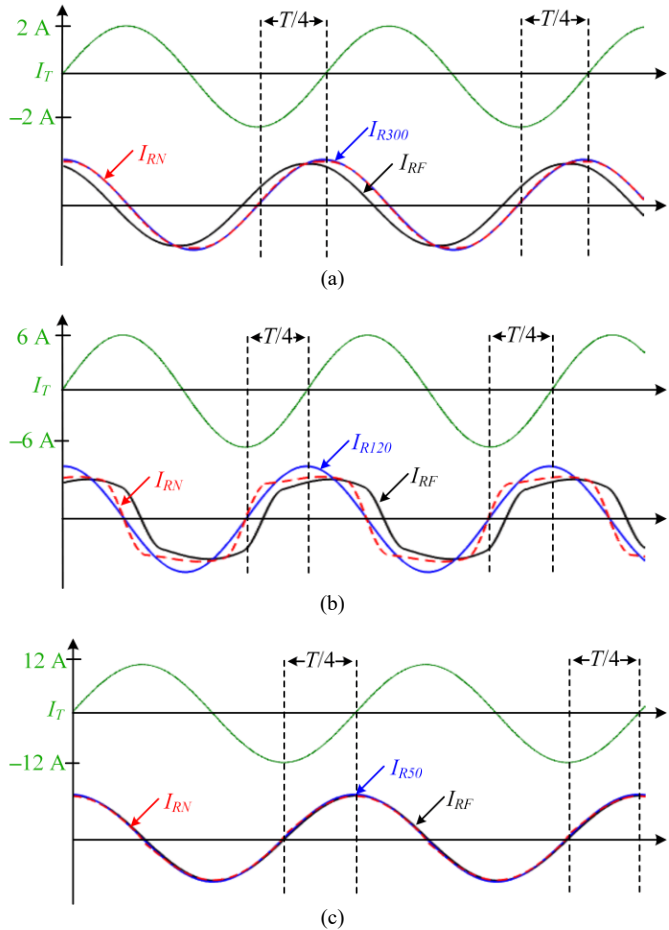

Fig. 7. Currents of the transceiver and the receivers at different frequencies. (a) At 300 kHz. (b) At 120 kHz. (c) At 50 kHz.

However, when the system operates at 50 kHz, the phase difference between $I_{RF}$ and the resonant receiver current $I_{R50}$ is negligible. Then, the tuning process is skipped, and the rated values of all $I_{RF}$, $I_{RN}$, and $I_{R50}$ are basically equal (Fig. 7(c)).

TABLE III
KEY EXPERIMENT PARAMETERS

| Item | Value/Type | Unit |
| --- | --- | --- |
| WPT frequency range ($f_T$) | 65–125 | kHz |
| Transmitter coil inductances ($L_T$) | 45 | µH |
| Receiver coil inductances ($L_R$, $L_{R75}$, $L_{R132}$) | 38, 38, 38 | µH |
| Mutual inductance ($M_R$) | 9 | µH |
| Compensation capacitances ($C_{R1}$, $C_{R2}$) | 22, 147 | nF |
| Load resistances ($R_L$, $R_{75}$, $R_{132}$) | 5, 5, 5 | Ω |
| Transmission distance | 20 | mm |
| Diameter of the transmitter coil | 300 | mm |
| Diameter of the receiver coil | 120 | mm |

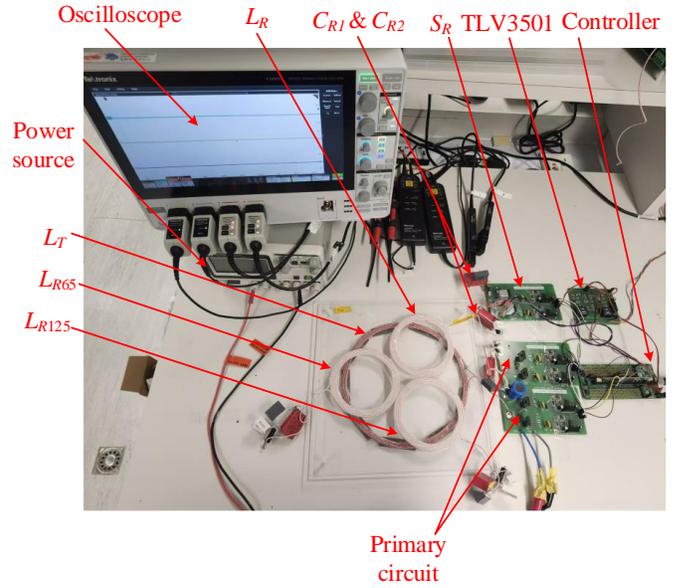

Fig. 8. Experimental setup.

## V. EXPERIMENTS

We implemented an experimental setup to demonstrate the performance of the new energy decryption system for wireless power hacking (Fig. 8). The experimental setup contains one transmitter and three receivers: The actively hacking thief $L_R$ and two receivers with fixed resonance, specifically receiver $L_{R65}$ resonant at 65 kHz and $L_{R125}$ resonant at 125 kHz. The hacking receiver uses $V_{CR1}$ as the control signal with a 0.75 µs filter time; the system can therefore harvest energy from 65 kHz to 125 kHz. Table III summarizes the key parameters.

This experiment uses a time-division switched capacitor array also on the transmitter side so that the transmission frequency $f_T$ can be tuned quite flexibly. However, the receiver does not know the key, i.e., the sequence of frequencies. Although our experimental setup uses a fully tunable transmitter, practical encryption systems may use other setups, including discrete-frequency, for higher efficiency or utilization.

In the test of Fig. 9, $f_T$ jumps between 65 kHz to 125 kHz periodically. Without any information, the hacking receiver $L_R$ follows this change closely. When $f_T$ jumps from 65 kHz to 125 kHz, the hacker only needs some 120 µs or 15 periods to detect the new frequency and control $S_R$ as well as tune the second capacitor $C_{R2}$ to a lower duty ratio. However, when the transmitter switches the operating frequency from 125 kHz to 65 kHz, the power thief needs about 200 µs or 15 periods to detect the change and adjust the compensation network. This rapid response outcompetes any previous attack by some three powers of ten.

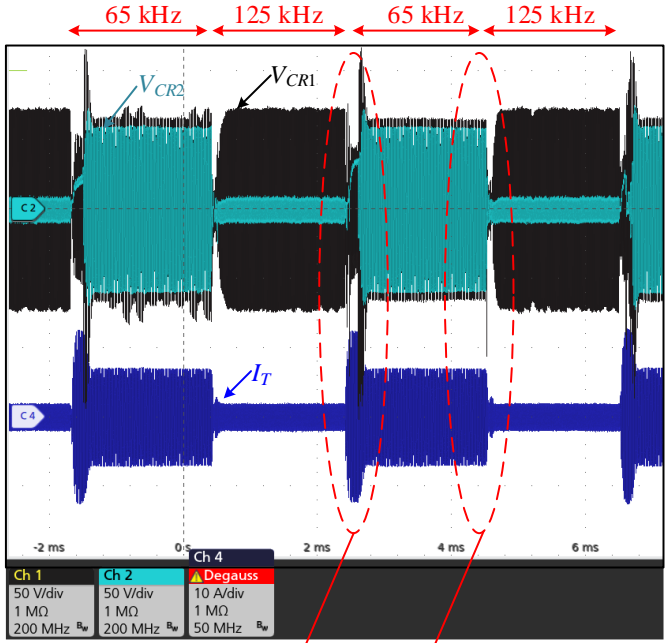

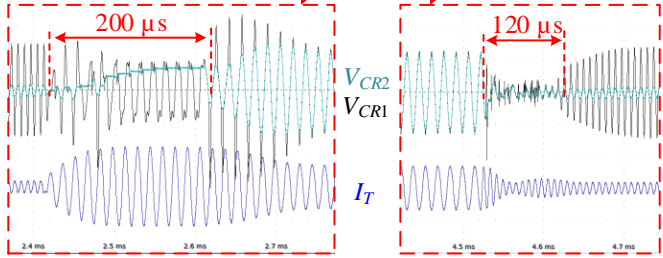

Fig. 9. Transmitter current and hacker compensation voltages at frequency transient points, where $I_T$ is the transmitter current, while $V_{CR1}$ and $V_{CR2}$ respectively are the voltages of the first and second compensation capacitors.

Therefore, the power stolen by the hacker can be near the fully resonant receiver's power within 200 μs. Fig. 10 compares the power with the fixed resonators, where $I_{R65}$ and $I_{R125}$ are the respective output load currents of the 65 kHz and 125 kHz receivers.

In addition to the rapid tracking dynamics, the presented system can furthermore extract a large share of the available power. The respective rated values of $I_R$, $I_{R65}$, and $I_{R125}$ are 2.75 A, 2.9 A, and 0.6 A (Fig. 11 (a)). Thus, the extracted power by the hacker, the fixed 65 kHz receiver, and the fixed 125 kHz receiver respectively is 37.8 W, 42 W, and 1.8 W. The ratio of them is 0.9:1:0.04. At high frequencies, the respective rated values of $I_R$, $I_{R65}$, and $I_{R125}$ are 1.4 A, 0.45 A, and 1.7 A (Fig. 11 (b)). Thus, the unauthorized hacking receiver can extract two-thirds of the power received by the authorized receiver in less than 200 μs or 15 to 20 cycles.

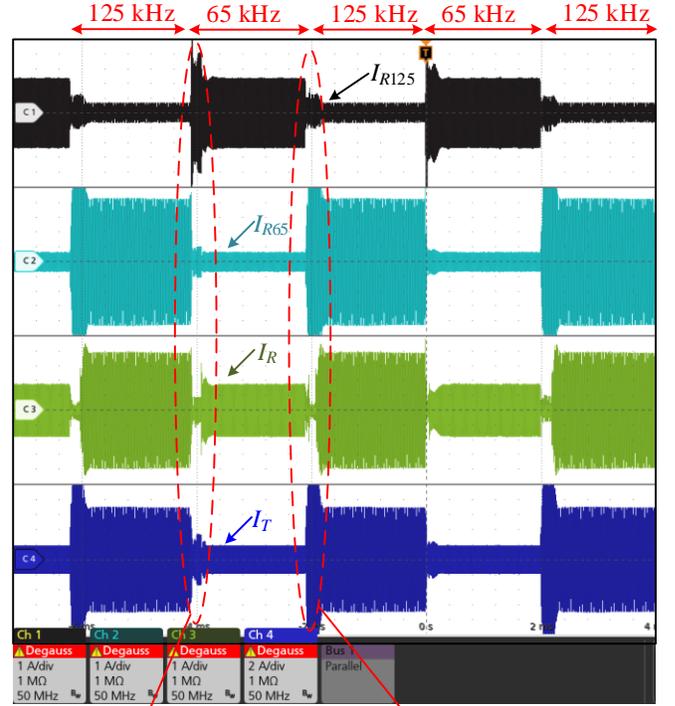

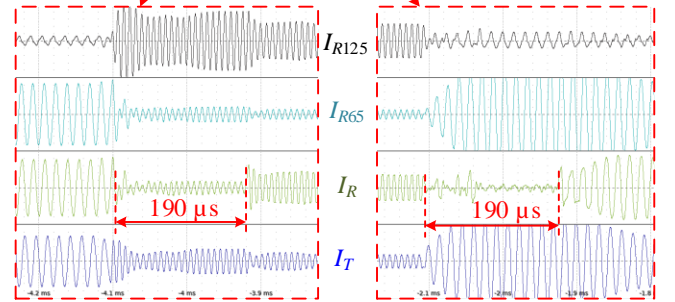

Fig. 10. Transmitter and load currents at frequency switching points, where $I_T$ is the transmitter current, while $I_R$, $I_{R125}$, and $I_{R65}$ respectively are the currents of the intruder, 125 kHz receiver load, and 65 kHz receiver load.

Furthermore, Fig. 11 clarifies that the fully resonant receiver's load currents are 90° ahead of the transmitter current $I_T$, which is consistent with the theoretical analysis based on (3). The hacker calibrates the $T_{ON}/T_{OFF}$ times by detecting the upward zero-crossings of the load voltage and the transmitter capacitor voltage, which should have a 180° difference under ideal conditions.

Moreover, Fig. 12 illustrates the relationship between switching on/off time (phase shift of $I_R$) and hacking performance. As the system is operating at 65 kHz (low frequency), $T_{OFF}$ should be very small to increase the equivalent capacitance $C_{RE}$. However, if we increase $T_{OFF}$ on purpose, the hacker is not resonant anymore, so the phase difference between $I_T$ and $I_R$ increases and the value of $I_R$ drops accordingly.

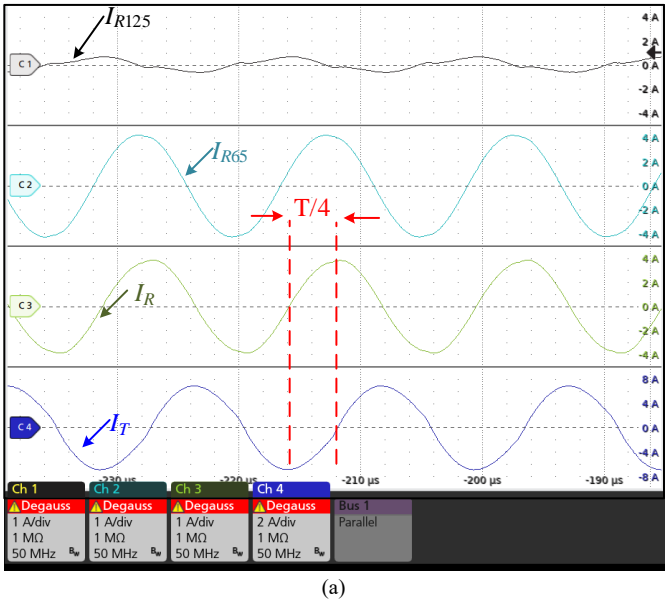

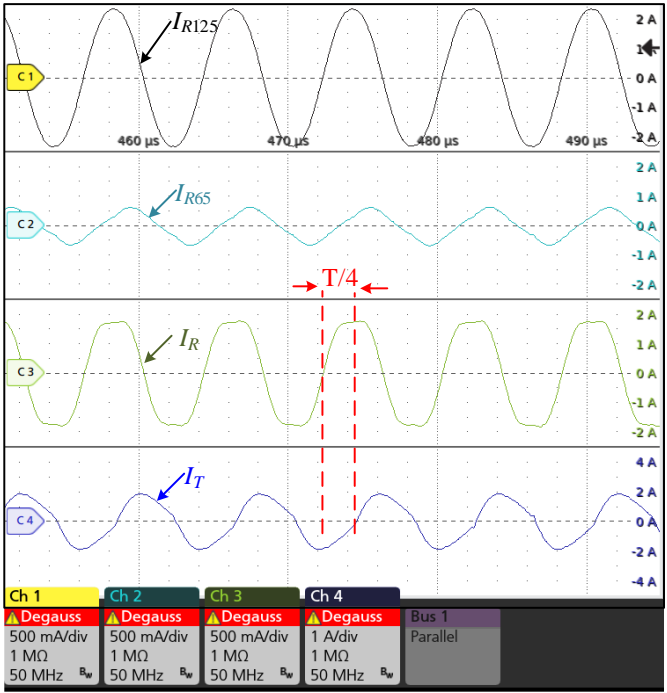

Fig. 11. Transmitter current and load currents under different frequencies, where $I_T$ is the transmitter current, while $I_R$, $I_{R125}$, and $I_{R65}$ respectively are the currents of the intruder, 125 kHz receiver load, and 65 kHz receiver load. (a) For a transmitter frequency of 65 kHz. (b) For a transmitter frequency of 125 kHz.

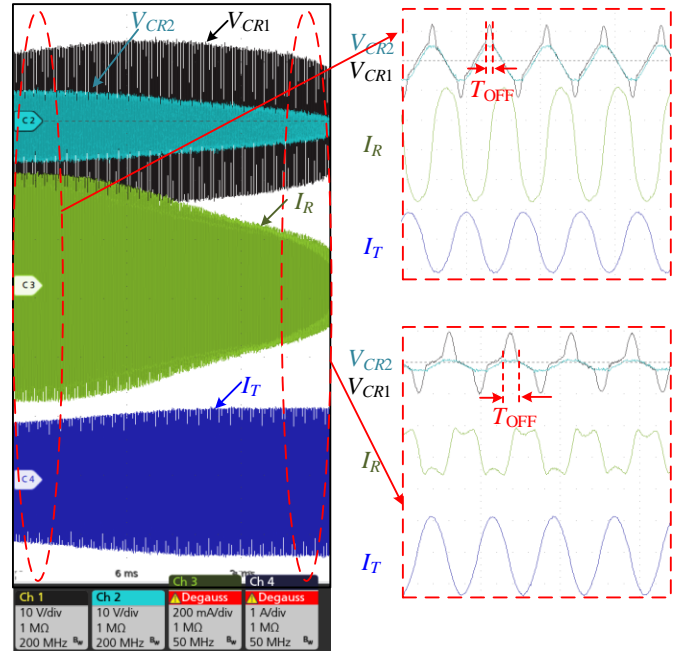

Fig. 12. Relationship between switching on/off time (receiver current phase shift) and hacking performance at 65 kHz.

## VI. Conclusion

This paper presented a rapid attack on encrypted wireless power transfer. The system can track the changing frequency and impedance in about 15 to 20 cycles and is therefore not much slower than a passive resonator. It is by several powers of ten faster than previous attacks and may render reinforcement of wireless power encryption rather challenging.

The key hardware element that enables this hacking performance is a tunable resonator, which adjusts the capacitance with time-division switching. The key to control optimization is the exploitation of the phase difference between the transmitter and the interceptor to regulate the resonator in advance, then looking up the table in real applications. Thus, the new system is a microsecond-fast energy hacker, and it was found to be substantially faster than any maximum-receiver-current regulation or tracking through an ADC, as suggested earlier. In real applications, 200 µs is typically substantially shorter than the time for an EV to even pass through a 20-cm coil (360 µs for as high as 200 kph). Furthermore, to put the timing into perspective, the less than 200 µs initialization and synchronization compare well with, for example, Bluetooth and even 5G, which can be used to exchange key sequences and synchronize authorized receivers through a parallel data channel. Considering some studies indicate a possibility to reduce latencies and synchronization of enhanced 5G and 6G to some 0.125 ms (in lab environments) [42, 43], future optimized attacks may do so too and shorten the regulation time to less than envisioned 6G latencies.

The presented system furthermore eliminates any additional sensor coil and instead uses the power receiver coil to sense the magnetic field. This simplification greatly reduces the cost and complexity of any hacking contraption. The extractable power, however, is not diminished and can still exceed two-thirds.

Important in that context is that any stolen energy, even only a tenth of that level would be a problem for an operator. For a hacker, the power is free so that efficiency beyond a certain point is also no issue, it would be at the expense of the operator anyway. Thus, the demonstrated power theft should be alarming.

Overall, in light of the recently rapidly spreading wireless power applications from initially toothbrushes via cellphones, consumer electronics, and household appliances to now cars, encryption techniques and the cybersecurity of wireless power may need more attention. This paper demonstrated that simple frequency-varying WPT encryption—no matter which cipher or sequence in the background—is no longer reliable. Perhaps the power supplier can use policy or legal means, such as building paid user lanes or adding monitors along the roads (as long as the policy states that the cameras will not violate privacy). Of course, we also hope that more scholars can devise better technical means to protect energy.

We are going to further optimize the attack. We see potential and intend to shorten the deciphering speed (maybe as fast as two to three cycles) and also reduce the latency to improve the hacking efficiency. Moreover, we aim at an intelligent interceptor that could harvest the maximum energy in multi-frequency magnetic fields and in the presence of distortion or jamming. Thus, it should be able to select the frequency with the highest power content and not be affected by other frequency content.